\documentclass[a4paper,11pt]{article}
\usepackage{pos}
\usepackage{braket}
\usepackage{graphicx}
\usepackage{cleveref}
\graphicspath{{./Figures/}}

\title{Interplay of Gauss Law and the Fermion Sign Problem in Quantum Link Models with Dynamical Matter}
\ShortTitle{Interplay of Gauss Law and the Fermion Sign Problem}

\author*[a,b]{Pallabi Dey}
\author[c]{Debasish Banerjee}
\author[d]{Emilie Huffman}
\affiliation[a]{Theory Division, Saha Institute of Nuclear Physics, 1/AF Bidhan Nagar, 
Kolkata 700064, India}

\affiliation[b]{Homi Bhabha National Institute, Training School Complex, Anushaktinagar, 
Mumbai 400094,India}

\affiliation[c]{School of Physics and Astronomy, University of Southampton, University Road, SO17 1BJ, UK}
\affiliation[d]{Department of Physics and Center for Functional Materials, 
Wake Forest University, Winston-Salem, North Carolina 27109, USA}

\emailAdd{pallabi.dey@saha.ac.in}
\emailAdd{D.Banerjee@soton.ac.uk}
\emailAdd{ehuffman@wfu.edu}

\abstract{Quantum Link Models with dynamical matter coupled to spin-$\frac{1}{2} \ \rm U(1)$ gauge fields in $d=2+1 $ and $3+1$ can potentially give rise to the Coulomb phase expected in quantum electrodynamics (QED) and other confining phases. Using exact diagonalization techniques, we show that the ground state in a class of models without the magnetic field always lies in the sector which satisfies $(G_e,G_o) = (d,\ -d)$, where $d$ is the spatial dimension and $e$ and $o$ are even and odd sites. It can be analytically proven that this sector is free of the fermion sign problem. We also demonstrate that a meron cluster algorithm for the problem naturally samples the ground states of the Hamiltonian in the aforementioned Gauss Law sector.}
\FullConference{The 42nd International Symposium on Lattice Field Theory (LATTICE2025)\\
2-8 November 2025\\
Tata Institute of Fundamental Research, Mumbai, India\\}

\begin{document}
\maketitle

\section{Introduction}
The sign problem is a major challenge in the numerical simulation of strongly interacting fermionic systems
using Monte Carlo methods \cite{Troyer2005}. 
The computation of physical observables for these methods require an importance sampling of all possible configurations. In the presence of fermions, configurations in the occupation number (Fock) basis can acquire negative weights due to the odd number of fermionic exchanges. These cancellations between positive and negative contributions in the computation of physical observables lead to an exponential growth of statistical errors with system size or inverse temperature, making simulations of large or low-temperature systems practically impossible.
A similar, more general phenomenon arises in models where the Boltzmann weights are non-positive or complex, for example, in systems with finite chemical potential \cite{Nagata2022}, model with topological-$\theta$ terms \cite{PhysRevLett.75.4524} or Hubbard model away from half-filling \cite{Arovas2022}. In all these cases, cancellations between contributions
with positive and negative weights create a severe sign problem.

There are multiple approaches to tackle sign problem \cite{Wu2005,Assaad2007,Chandrasekharan2010,Cristoforetti2012,Huffman2014,Huffman2016,Huffman2017,
Gattringer:2016kco,Wang2016,Li2019,Gantgen:2023byf}, but in particular there is a class of problems which can be simulated by the Meron Cluster algorithm \cite{Chandrasekharan1999, Chandrasekharan_2000, Liu2020}.
The key idea of the method is to first perform an analytic summation of certain fermionic world-line configurations to cancel those with equal and opposite signs, and subsequently to completely avoid generating such configurations during the importance sampling.  It was first developed for the $2d \ O(3)$ sigma model \cite{PhysRevLett.75.4524} at $ \theta = \pi $, where merons act as half-instantons, and since has been applied to four-fermion Hamiltonians and free fermion systems in certain parameter ranges \cite{Chandrasekharan1999, Chandrasekharan_2003, Chandrasekharan_2000}.

Subsequently, additional meron-cluster amenable models which incorporate gauge fields coupled to fermions have been identified as a promising next step \cite{Banerjee:2023cvs, PintoBarros:2024oph}, as they allow the study of phase diagrams for models relevant to lattice QCD. Another key motivation is that these models  can be realized in analog and digital quantum simulator experiments which can simulate constrained lattice gauge theories and spin models in $d = 1, \ 2$ \cite{Martinez2016,Bernien2017,Ebadi2020,Yang2020,Zhou2021,Semeghini2021,Moss2023,Gonzalez-Cuadra2024,
Schuhmacher2025}.

Here, we study such a model: spin-$\frac{1}{2} \ \rm U(1)$ gauge fields coupled to single flavor fermions in $d = 2, \ 3$ spatial dimensions. The $\rm U(1)$ gauge variables on the bonds are represented as quantum links, which realise gauge invariance with finite dimensional Hilbert spaces. In particular, spin $s = \frac{1}{2}$ links entail a local two-dimensional Hilbert space.
Using this model, we explore how Gauss law constraints order different superselection sectors and explore the role of the fermion sign problem in them. Our analysis combines exact diagonalization with the meron-cluster algorithm and our results guide the application of quantum simulators to solve sign problems. This approach allows us to study strongly interacting fermions coupled to gauge fields in a controlled and computationally efficient way.

\section{Model with Spin-$\frac{1}{2} \ \mathrm{U(1)}$ Gauge Links Coupled to Matter}
We consider a lattice model of spin-$\frac{1}{2}$ $\mathrm{U(1)}$ gauge fields coupled to fermionic matter in $d$ spatial dimensions with lattice volume $\cal V$, described by the Hamiltonian:
\begin{figure}[!tbh]
    \centering
    \includegraphics[width=0.65\linewidth]{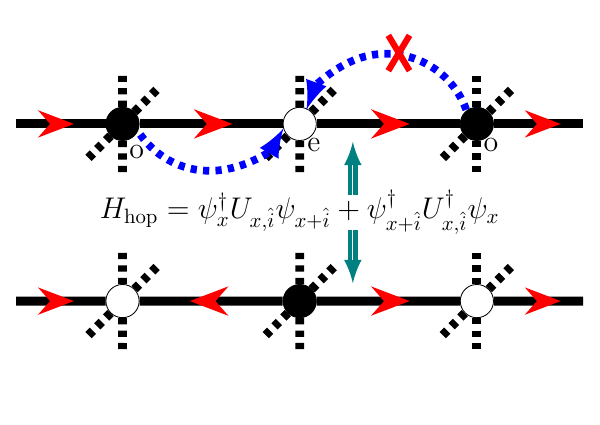}
    \caption{The fermion hop from left to right is accompanied by a $\sigma^- (U^\dagger)$ operator on the link, causing the spin-$1/2$ electric flux to flip its orientation, while the right to left hop is accompanied by a $\sigma^+ (U)$ operator on the link and is thus constrained by the state of the flux on the link.
    }
    \label{fig:H_hop}
\end{figure}
\begin{equation}\label{eq:H_U1}
\begin{aligned}
 H &= -t\sum_{x,i} \left[ \left(\psi^\dagger_x U_{x, \hat i} \psi_{x+\hat i} 
    + {\rm h.c.} \right) 
    -  E_{x, \hat i}\left(n_{x+\hat i} - n_x \right) \right] + (V-t) \sum_{x,i}  \left(n_x - \frac{1}{2}\right)\left(n_{x+\hat i} 
    - \frac{1}{2}\right) - \frac{t \cdot d \cdot \cal{V}}{4}.
\end{aligned}
\end{equation}
The first term corresponds to the hopping term where the hopping of the spinless fermions between the nearest neighbors are constrained with the gauge fields $\rm{U_{x,i}} (\rm{U^\dagger_{x,i}})$ creating (annihilating) the electric fluxes, represented by spin$\frac{1}{2}$ operators. Thus, the hopping of fermions is constrained by the gauge links, as depicted in \cref{fig:H_hop}.
The second term is a designer term that makes the meron-cluster algorithm applicable for the Hamiltonian \cite{Banerjee:2023cvs}. The third term is the four fermi interaction between neighboring sites. This model has no sign problem for the meron method when $V\geq 2t$, as discussed later.

This model exhibits a set of local symmetries as the Hamiltonian commutes with the Gauss law operator ${G_x}$ at all values of $x$. For the $\rm U(1)$ theory, the local Gauss law operator is defined as,
\begin{equation} \label{eq:GL}
    G_x = n_x + \left( \frac{(-1)^x - 1}{2} \right) - \sum_{i} (E_{x,\hat i} - E_{x-\hat i,\hat i}), 
\end{equation}
where $n_x$ denotes matter occupation number and the second term represents a staggered background charge that is essential for maintaining global charge conjugation symmetry and ensuring a charge-neutral vacuum. The final term corresponds to the lattice divergence of the electric field. 
Because of $[G_x, H] = 0$, it leads to the fragmentation of the Hilbert space into different superselection sectors, labeled by the eigenvalues of $G_x$. 
To discuss the discrete symmetry in those superselection sectors, we will use the notation $(Ge, \ Go)$ for the even and odd-site GL sectors.

In this work, we are mainly focused on the shift symmetry $\mathcal{S}_k$, which is a discrete translation of all fields by one 
lattice unit along the $k$-th spatial direction. This is the discrete chiral symmetry
for this model. The symmetry is imposed on the operators as:
\begin{equation}\label{eq:shiftsym}
    \begin{aligned}
       {}^{\mathcal{S}^{k}}\!\psi_x &= \psi_{x+\hat k}, 
        {}^{\mathcal{S}^{k}}\!{{U_{x,\hat i}}} = U_{x+\hat k, \hat i}, 
        {}^{\mathcal{S}^{k}}\!{{E_{x,\hat i}}}  = E_{x+\hat k, \hat i}.        
    \end{aligned}
\end{equation}
However, under the action of \cref{eq:shiftsym}, the Gauss law operator at site $x$ transforms as, 
\begin{equation}
    \begin{aligned}
        G_x \xrightarrow{\mathcal{S}^k} G_{x+\hat{k}} + (-1)^x.
    \end{aligned}
\end{equation}
Under this transformation, each basis state in the GL sectors $(G_e, \ G_o) = (0, \ 0), \ (2,\ -2), \ (3, \ -3)$ is mapped to a unique basis state of the shift symmetric GL sectors $(1, \ -1), \ (-1, \ 1), \ (-2, \ 2)$ respectively, which makes each GL sector isomorphic to its corresponding shift GL sector, independent of spatial dimensions \cite{dey2025fermionsignproblemgauss}.

\section{Sign Problem in different sectors}
In this section, we will explore the Gauss law sectors analytically in $d=3$ dimensions. The local Hilbert space in the GL sector $(3,\ -3)$ is constructed by solving the equation $n-\nabla . E = 3$ and  $n-1-\nabla . E = -3$ for even and odd sites respectively. 
In $d=3$, the first equation gives $ 1 \ (6)$ solution(s) for $n = 0 \ (1)$ on even sites, while the second equation gives $6 \ (1)$ solution(s) for $n = 0 \ (1)$ on odd sites, resulting $ (7, \ 7)$ allowed configurations for (even, odd) sites. In the shifted GL sector 
$(-2, \ 2)$, the Gauss law equations on even and odd sites are swapped relative to the sector
$(3, \ -3)$, which leads to the same number of solutions.
With the same counting procedure for the GL sector $(0,0)$ and its shifted partner $(1,-1)$, we find $(35,35)$ allowed configurations on (even, odd) sites. For the GL sector $(2, \ -2)$ (and its shift partner $(-1, \ 1)$), there are $(21, 21)$ allowed configurations on (even, odd) sites. Also, there is one GL sector $(4, \ -4)$ (and its shifted partner $(-3, \ 3)$) where there is a single state of completely frozen fermions which do not hop at all.

\begin{figure*}[!tbh]
  \centering
  \begin{minipage}[t]{0.48\textwidth}
    \vspace{0pt}
    \centering    \includegraphics[width=\linewidth,keepaspectratio=false]{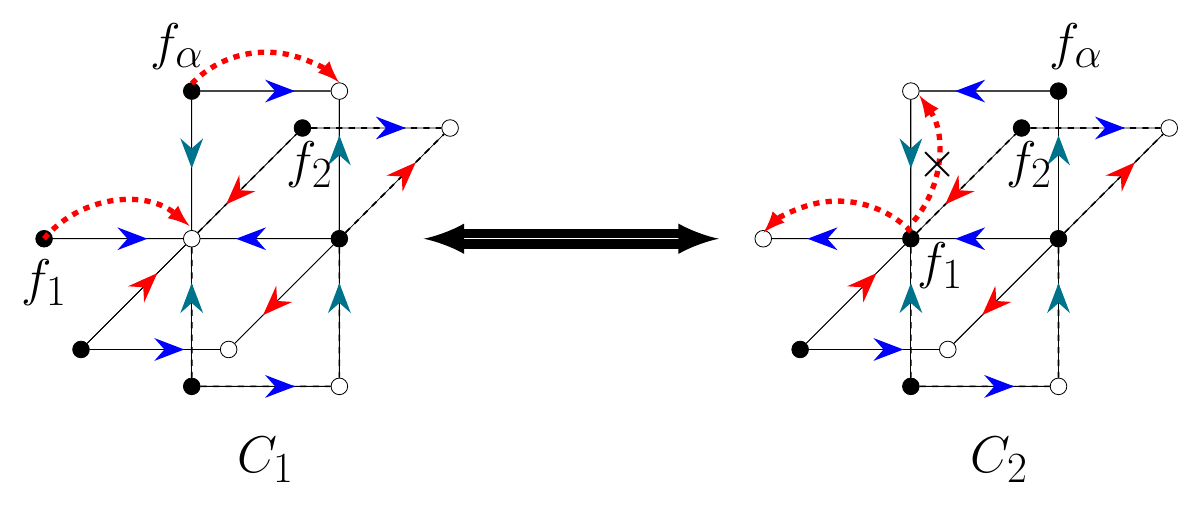}
  \end{minipage}%
  \hfill
  \begin{minipage}[t]{0.5\textwidth}
    \vspace{0pt}
    \centering
    \includegraphics[width=\linewidth]{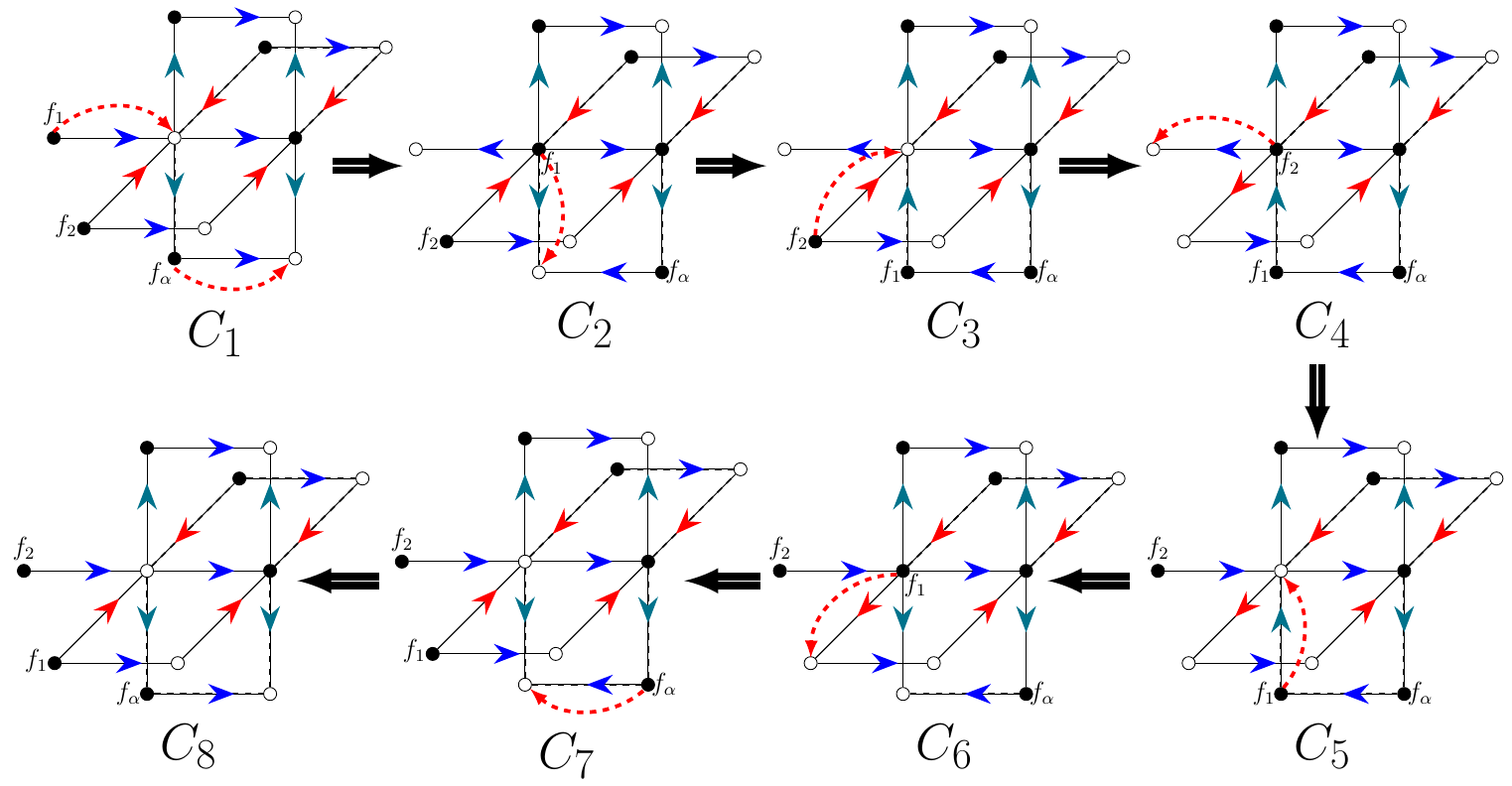}
  \end{minipage}
  \caption{(Left): The orientation of the gauge links in the sector $(3,-3)$ in $d=3$ do not allow the movement of fermions beyond one lattice spacing. Thus, positions of $f_1$ and $f_2$ cannot be switched by the action of the Hamiltonian. (Right) The GL constraints in the sector $(0,0)$ in $d=3$ 
  are relaxed enough to allow fermions $f_1$ and $f_2$ to exchange positions with each 
  other following the 
  general prescription described in the text.}
  \label{fig:GLd3}
\end{figure*}

To describe the sign problem explicitly, the analytic proof proceeds by identifying two fermions in a spatial configuration and interchanging 
their positions by acting with the Hamiltonian, keeping the positions of all other fermions fixed. 
We now focus on the $(3,-3)$ sector in $d=3$. 
Among the $(7,7)$ allowed solutions, a representative configuration is shown in \cref{fig:GLd3}. We consider $(f_1,f_2)$ as two fermions with 
positions $(r_1,r_2)$, and we also identify an auxiliary fermion $f_\alpha$ at position 
$r_\alpha$ which we will use to facilitate the interchange in positions of $(f_1,f_2)$ through the action of the Hamiltonian in \cref{eq:H_U1}.
The procedure is to move $f_1$ to $r_\alpha$ by temporarily displacing $f_\alpha$, while $f_2$ is moved to 
$r_1$. It is then completed by pushing $f_1$ to $r_2$, and $f_\alpha$ is returned to its original 
location. 
As shown in \cref{fig:GLd3} (left), just after a single hop of $f_1$ by the action of $\psi^\dagger_{r_1} U_{r_1, \hat i} \psi_{r_1+\hat i}$, we encounter links 
(fixed by GL) which forbid further hopping of $f_1$ --- only the reverse hop back to $r_1$ 
is allowed. Although the process is shown for one example, we have verified
this to be the case for \emph{all} configurations locally in the GL sector $(3,-3)$
(and by extension to its shift partner).

This is in contrast to the case of the sector $(0, 0)$, which is a relevant sector in particle physics, shown
in \cref{fig:GLd3} (right), where the above procedure works. Thus, configurations $C_1$ and $C_8$
differ by an overall sign when computations are performed in this basis. It results in a severe 
sign problem in a QMC sampling, which can be potentially solved with the meron concept.
Moreover, this argument generalizes to lower dimensions. In $d=2$ the sector $(2,-2)$ has no sign problem, but the $(0,0)$ has one \cite{dey2025fermionsignproblemgauss}. This 
analytical argument also explains why no sign problem was found in $d=1$ in \cite{Banerjee:2023cvs}.

\section{Algorithm}
\begin{figure}
    \centering
    \includegraphics[scale=0.7]{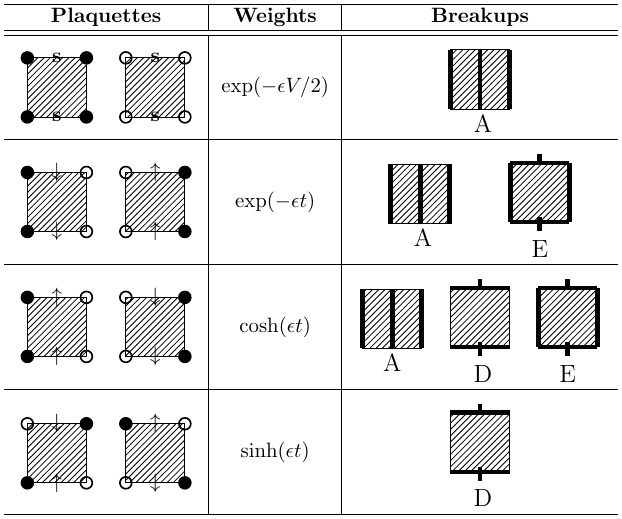}
    \caption{Breakup and corresponding weights for the spin-$\frac{1}{2}$ U(1) gauge links 
    coupled with matter Hamiltonian in \cref{eq:H_U1}.}
    \label{fig:breakup}
\end{figure}

The algorithm is constructed in terms of worldline configurations in the fermion occupation-number basis ($n_x$) and the electric flux basis ($E = s^3$) for the gauge links. Derivation of this algorithm begins with a Suzuki–Trotter decomposition of the Hamiltonian, leading to a discrete Euclidean-time representation.
The Hamiltonian is decomposed into sets of mutually commuting operators in $d$ spatial dimensions,
\begin{equation}
    H = \sum_i^{2d} H_i,
\end{equation}
with 
\begin{equation}
    \begin{aligned}
        H_i &= \sum_{\substack{x= (x_1, x_2, x_3,\cdots, x_d) \\ x_i \in \text{even}}} h_{x,i}, &\quad
        H_{i+d} &= \sum_{\substack{x = (x_1, x_2, x_3, \cdots, x_d) \\ x_i \in \text{odd}}} h_{x,i}.
    \end{aligned}
\end{equation}
The Suzuki-Trotter decomposition \cite{Chandrasekharan_2000, Banerjee:2021zed} allows the Boltzmann operator $e^{-\beta H}$ to be expanded over
$2dN_t = \beta$ discrete Euclidean time slices of spacing $\epsilon = \frac{\beta}{N_t}$.
The partition function in $d=3$ dimensions is then given by,
\begin{equation}{\label{eq:Z}}
\begin{aligned}
   Z &= \mathrm{Tr}(e^{-\beta H}) = \sum_{\{s, n\}}
     \prod_{\tau}
     \bra{ s(x,\tau), n(x,\tau)}  e^{-\epsilon H_1} \cdots e^{-\epsilon H_6} 
     \ket{ s(x,\tau-1), n(x,\tau-1)},
\end{aligned}
\end{equation}
with periodic boundary conditions in Euclidean time, $\ket{ s(x,0), n(x,0)}=\ket{ s(x,\beta), n(x,\beta)}$.
In this representation, the occupied sites define the fermion worldline that are closed in Euclidean time, and electric flux variables track the gauge field dynamics. Each Trotter slice encodes a local configuration of fermion occupation and link variable, and the sequential operation of the transfer matrix, $e^{-\epsilon H_i}$, generates the full worldline ensemble. These worldline configurations form the basis for constructing clusters. Evaluating the matrix elements of the
transfer matrix allows the partition function to be expressed in terms of a local action and sign factor. The contribution from each plaquette can be written as,
\begin{align}
W_{\rm plaq} &= e^{-S[n(x,\tau),n(x+i,\tau),n(x,\tau+1),n(x+i,\tau+1); s^3(x,\tau),s^3(x,\tau+1)]},\\
  Z            &= \prod_{x,\tau} \sum_{n(x,\tau),s^3(x,\tau)} {\rm sign \left[ \{n \}\right] }~ W_{\rm plaq},
\end{align}
where, the local action $S[n(x,\tau),n(x+i,\tau),n(x,\tau+1),n(x+i,\tau+1);s^3(x,\tau),s^3(x,\tau+1)]$ couples the fermion occupations on neighbour sites at time slice $\tau$ with their neighbours in $\tau+1$, as well as the corresponding electric flux variables on these time slices, while ${\rm sign \left[ \{n \}\right] }$ keeps track of the negative signs arising from fermionic worldlines.
The weight of plaquette is computed from the local matrix elements 
$\braket{s_b, n_b | e^{-\epsilon H_b}| s'_b, n'_b }$, where $b$ is a nearest-neighbor bond, 
$b = {\{x, x+i}\}$. By introducing appropriate breakups associated with each plaquette, 
the matrix element products from \cref{eq:Z} can be decomposed into clusters, which preserve the allowed 
local configurations while maintaining detailed balance. Each active plaquette (shaded in gray) must correspond to one of the configurations listed 
in \cref{fig:breakup}. 

The term "breakup" refers to a specific way of connecting a subset of the six degrees of freedom in an active plaquette (four fermions at the corners and two gauge fluxes on the joining bonds) such that detailed balance is satisfied while these degrees of freedom are updated simultaneously. Different types of breakups connect different subsets of sites and links. For example, to satisfy detailed balance, the A, D, and E breakups shown in \cref{fig:breakup} are applied with probabilities determined by \cref{eq:detailedbalance} for $V \geq 2t$. Once breakups are applied at all active plaquettes in a configuration, one can define a cluster as the set of all fermions and gauge links which are connected together. Meron clusters, or merons, are those which on flipping changes the sign of weight of the configurations.

In this formulation, the link variables also contribute, either as additional lines in the 
A breakups or as binding extensions in the D and E breakups, representing a key generalization 
of the original meron cluster algorithm to gauge theories.
\begin{equation}\label{eq:detailedbalance}
\begin{aligned}
    W_A &= \exp(-\epsilon V/2) \\
    W_D &= \sinh(\epsilon t) \\
    W_{E} &= \exp(-\epsilon t) - \exp(-\epsilon V/2). 
\end{aligned}
\end{equation}
Flipping a cluster is defined as an operation that exchanges the occupied (up) and empty (down) 
sites (links). If a cluster is flipped, the magnitude of the weights of configuration remain 
the same. But if the cluster flip changes the sign of the configuration, it is a meron.
Rules for identifying a cluster as a meron are given by, 
\begin{equation}
\begin{aligned}
n_w + \frac{n_h}{2} &= \text{even}, & \quad & \text{odd \# of loops} \\
n_w + \frac{n_h}{2} &= \text{odd}, & \quad & \text{even \# of loops}.
\label{eq:meronrule2}
\end{aligned}
\end{equation} 
where, $n_w$ is the temporal winding number of the cluster, $n_h$ is the number of hops that the cluster makes to neighboring sites while encountering horizontal bonds (D-breakup or E-breakup).
With a certain choice of breakups it is possible to factorize the sign factor of a configuration 
$C$ into a product of the signs of each cluster, $\text{Sign}[C] = \prod_i^{N_c}\text{Sign}[C_i]$, 
where $C$ decomposes into $N_c$ clusters. By suitably flipping the clusters, one can reach the 
reference configuration. The reference configuration is defined by fermion worldlines only, arranged in a staggered (fermion) occupation pattern across the lattice. In this configuration, the fermions are stationary throughout the lattice and the Boltzmann weight of the configuration is positive, 
with $\text{Sign} = 1$. The existence of the reference configuration ensures the ergodicity and efficiency of the algorithm. The QMC update is as follows:
\begin{enumerate}
    \item Start from the reference configuration that contains only the fermion worldlines. 
    Choose an active plaquette randomly.
    \item If a random plaquette can switch to a different breakup, change it with a probability 
    based on its breakup weight.
    \item After a breakup is modified, re-examine the resulting configuration. If the modification 
    leads to merons in the configuration, then restore the breakup to its previous state and return 
    to step 2.
    \item For each cluster, flip all fermion occupations and flux variables with probability $1/2$.
\end{enumerate}

\section{Numerical Results}
We have investigated the model in both $d=2, \ 3$ spatial dimensions using both ED and the meron cluster algorithm. Here we address the important question of which GL sector determines the ground state in the low-temperature limit. At zero temperature, the energetically favored sector dominates. For $V/t > 0$, the Hamiltonian in \cref{eq:H_U1} selects the sector $(d,-d)$ and its shifted partner $(-d+1,\ d-1)$ as the ground state. In this sector, fermions are effectively localized, although local hopping by a single site is allowed, they cannot propagate far. However, at finite temperature, matter-anti-matter pairs can be created
as annealed disorder, and thus thermal fluctuations can connect different GL sectors.

\begin{figure}[!tbh]
    \centering
  \begin{minipage}[t]{0.5\textwidth}
    \vspace{0pt}
    \centering    
    \includegraphics[width=1.0\linewidth]{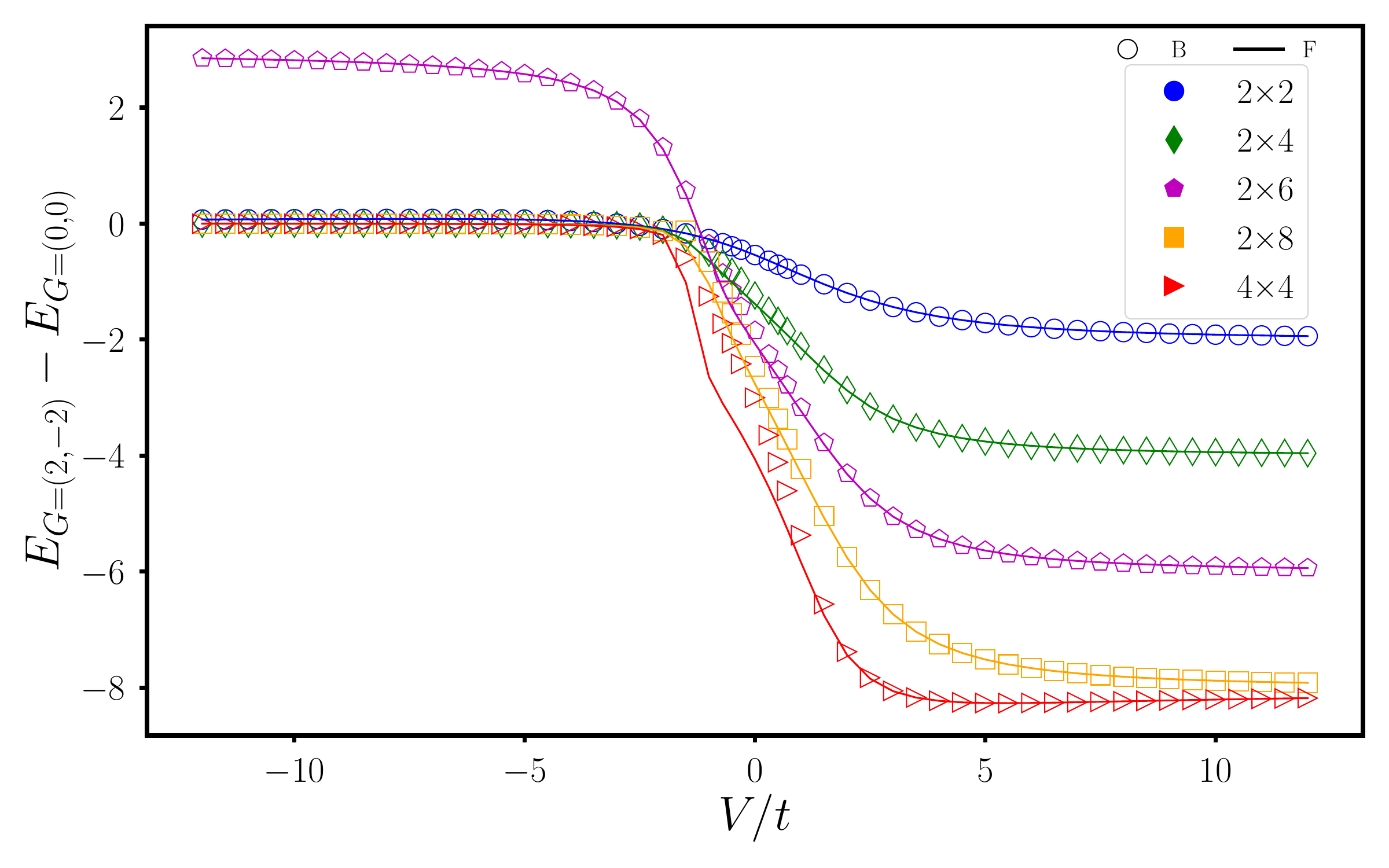}
  \end{minipage}%
  \hfill
  \begin{minipage}[t]{0.5\textwidth}
    \vspace{0pt}
    \centering
    \includegraphics[width=1.0\linewidth, keepaspectratio=false]{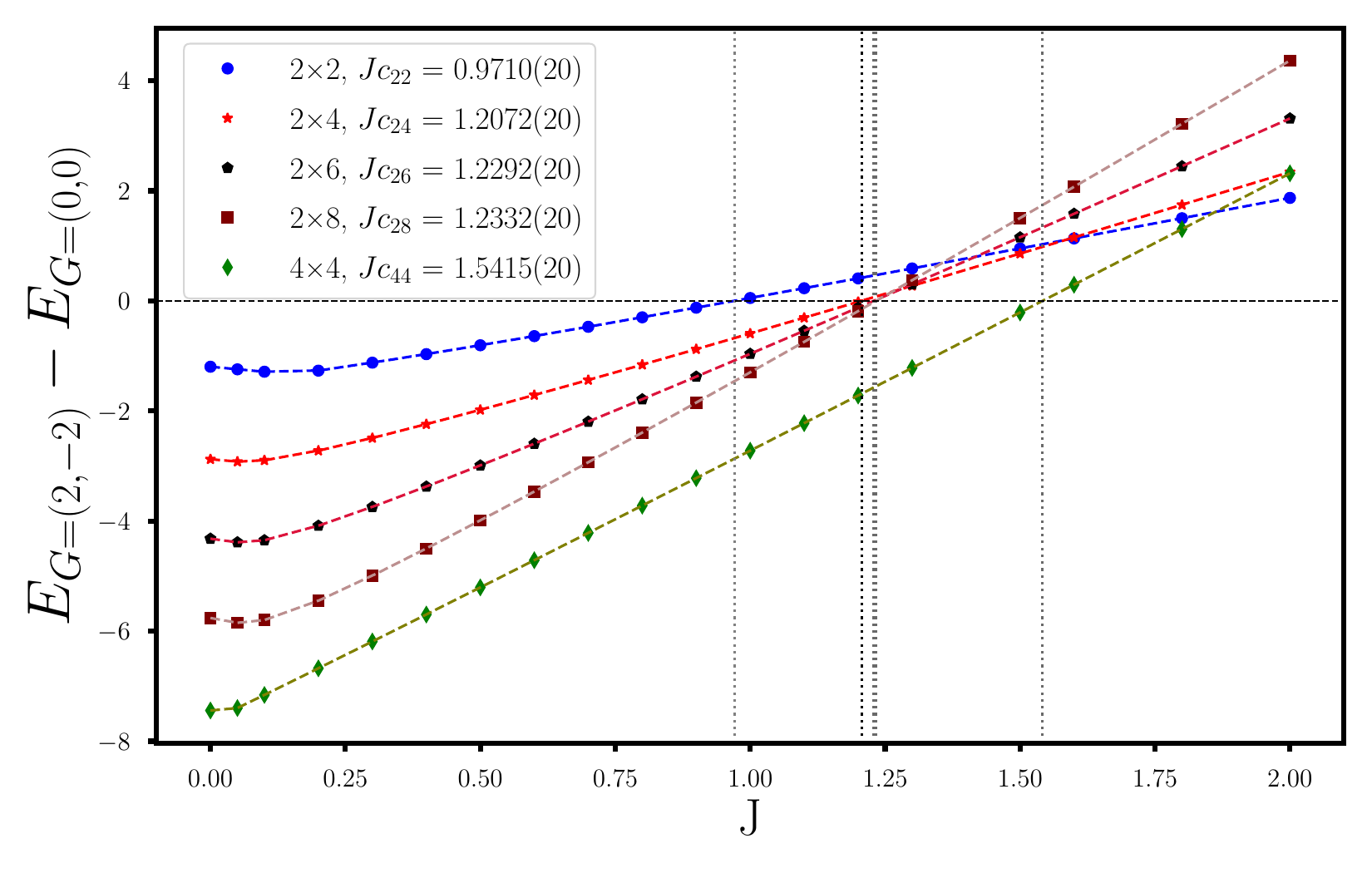}
  \end{minipage}
    \caption{(Left) Ground state energy difference between the GL sectors $(2,-2)$ and $(0,0)$
    for bosons (open symbols) and fermions (solid lines) respectively in $d=2$. While the fermionic 
    and bosonic results are the same for the sector $(2,-2)$, there is a difference in the 
    $(0,0)$ sector for $V/t \sim 0$, leading to the deviation between the two sets of data 
    in that region. (Right) Transition between the $(2,-2)$ and $(0,0)$ GL sectors as the magnetic coupling is increased.}
    \label{fig:delE}
\end{figure}

In $d = 2$, using large-scale ED, we study the ground
state at different $V /t $ regimes for both (hardcore) bosonic and fermionic matter up to 48 DOF (16 matter sites and 32 gauge links). Fig \ref{fig:delE} (left) shows the difference in ground-state energies between the two lowest sectors, $E_0^{(2,-2)} - E_0^{(0,0)}$, for both bosonic and fermionic matter (all other sectors lie higher in energy). For large $|V/t|$, local occupation numbers are effectively fixed, that makes the particle statistics whether bosonic or fermionic, irrelevant. In the regime $V/t \gg 0$, the four-fermion interaction dominates, leading to a charge-density-wave (CDW) phase. In this phase two CDW vacua belong to two different GL sectors, with equal and opposite (non-zero) values of the chiral condensate. For $V /t \ll 0$, both sectors have phase separated ground states featuring neighbouring sites that are both empty on one side of the lattice, and both filled on the other. 
There is a clear distinction between bosonic and fermionic matter in the regime where hopping processes are significant, particularly around $V/t \sim 0$, where particle exchange statistics become relevant. Detailed description of these phases can be found in our manuscript \cite{dey2025fermionsignproblemgauss}. 

\begin{figure}[!tbh]
    \centering
  \begin{minipage}[t]{0.5\textwidth}
    \vspace{0pt}
    \centering    
    \includegraphics[width=0.96\linewidth]{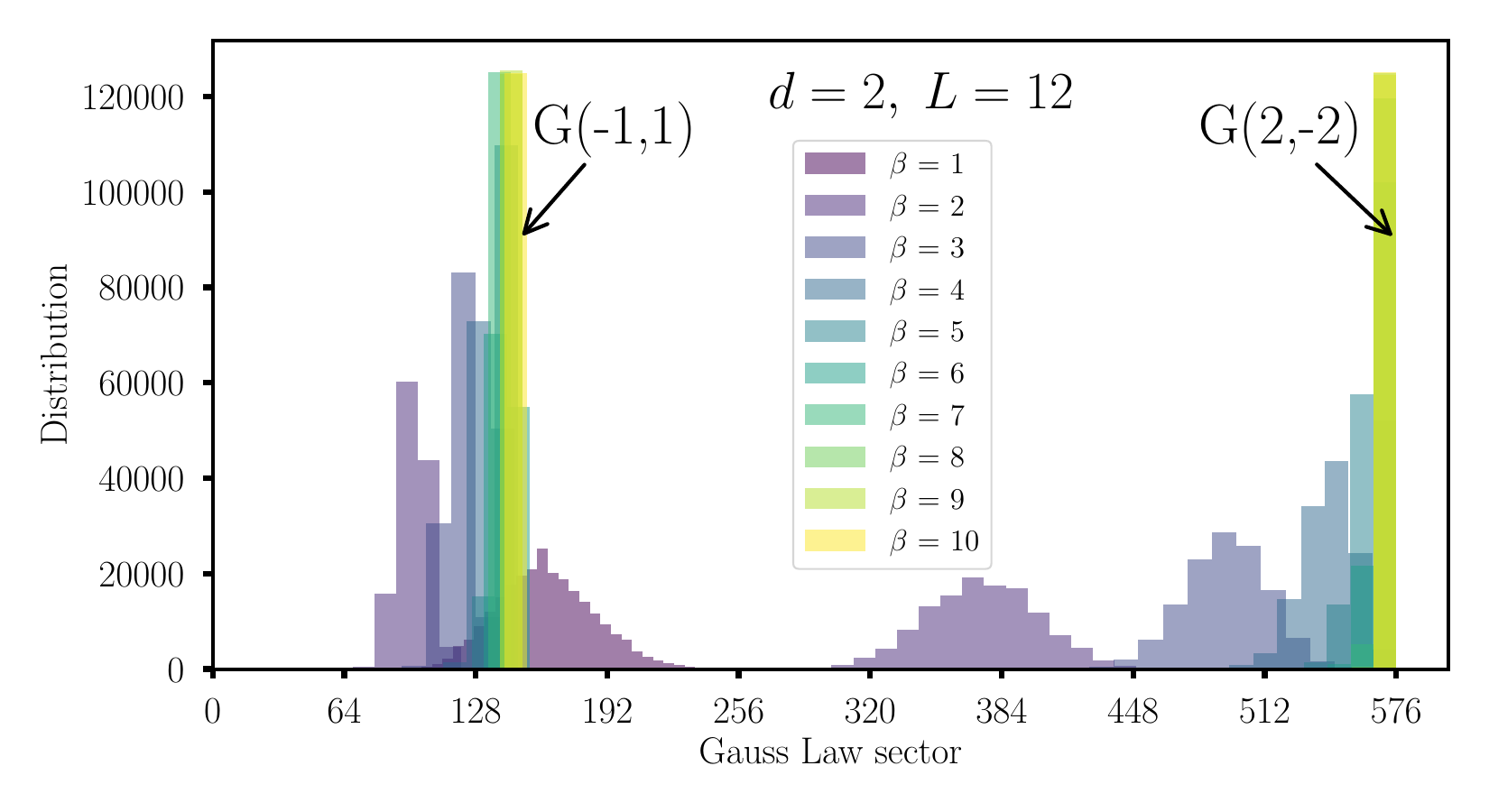}
  \end{minipage}%
  \hfill
  \begin{minipage}[t]{0.5\textwidth}
    \vspace{0pt}
    \centering
    \includegraphics[width=0.92\linewidth, keepaspectratio=false]{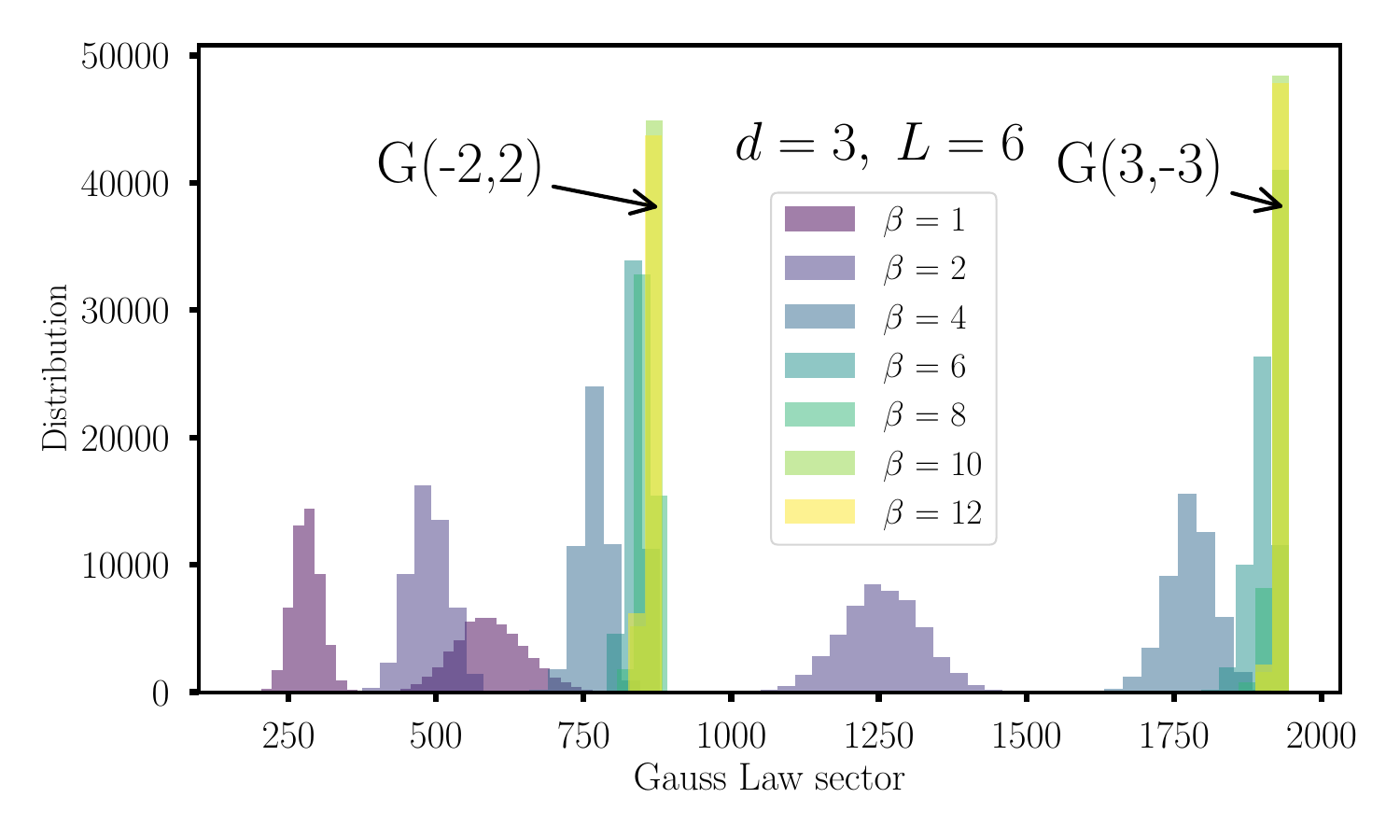}
  \end{minipage}
    \caption{The different GL sectors that are
sampled by the QMC algorithm at different $\beta$ in both $d=2$ and $d=3$ spatial dimensions. For low temperature (large $\beta$) only the GL $(d,\ -d)$ and its shifted partner arises.}
    \label{fig:GL3d_L6}
\end{figure}

We have benchmarked our ED results with QMC in both $d=2, \ 3$ spatial dimensions.
In \cref{fig:GL3d_L6}, we show how the distribution of different GL sectors changes with temperature in both dimensions. As the temperature decreases, the sector $(d,-d)$ and its shifted GL partner become dominant in both cases, indicating that these sectors control the low-temperature behavior.

Further we have investigated the role of including the magnetic energy term,
$-J \sum_\square (U_\square + U^\dagger_\square)$, to the Hamiltonian in \cref{eq:H_U1} in $d=2$ spatial dimensions. This analysis is performed for fermionic matter at $V=2t$. 
Fig \ref{fig:delE} (right) shows that there is transition from  the $(2, \ -2)$ sector to the $(0,0)$ sector as the magnetic term increases (for couplings beyond the critical value $J_c$).
For ladder lattices, we estimate $J_c \approx 1.23(1)$ shown as dashed vertical line 
in \cref{fig:delE} (right). For square lattices, $J_c$ is larger, but we expect it to be still 
$O(1)$ based on the scaling for $2\times2$ and $4\times 4$ systems.
The $6\times6$ lattice, with 108 degrees of freedom, is beyond the reach of our ED. The meron
algorithm does not yet work for $J \neq 0$.

\section{Conclusion}
In this work, we studied the fermion sign problem in different GL sectors of $\rm U(1)$ gauge theories where the gauge degree of freedom is represented using quantum links. While we focused on spin-$\frac{1}{2}$ representation of the gauge links, the approach developed here can be directly extended to higher spin-$S$ link variables as well as to truncated Kogut–Susskind Hamiltonians. The algorithm constructed in this work operates in the ground-state GL sector without relying on the meron concept, providing an efficient way to study this sector free of sign problems. An important next step is to explore whether the meron idea can be systematically implemented to eliminate the sign problem in other specific GL sectors, such as the $(0,0)$ sector, which will be 
important to explore the known and the exotic phases in non-perturbative version of quantum electrodynamics.


\section{Acknowledgements}
 We thank computing resources of SINP and DiRAC (computational facility for STFC) essential to derive these results.

\bibliographystyle{JHEP}
\bibliography{ref}

\end{document}